# Polyhedral units and network connectivity in calcium aluminosilicate glasses from high-energy x-ray diffraction

V. Petkov[1], S.J.L. Billinge[1], S.D. Shastri[2] and B. Himmel[3]

[1]*Department of Physics and Astronomy and Center for Fundamental Materials Research, Michigan State University, East Lansing, MI 48823, USA*

[2]*Advanced Photon Source, Argonne National Laboratory, Argonne, IL-60439, USA*

[3]*FB Elektrotechnik and Informationtechnik, Universität Rostock, Rostock, Germany*

Structure factors for $Ca_{x/2}Al_xSi_{1-x}O_2$ glasses ($x$=0,0.25,0.5,0.67) extended to a wave vector of magnitude Q= 40 Å$^{-1}$ have been obtained by high-energy x-ray diffraction. For the first time, it is possible to resolve the contributions of Si-O, Al-O and Ca-O coordination polyhedra to the experimental atomic pair distribution functions (PDF). It has been found that both Si and Al are four-fold coordinated and so participate in a continuous tetrahedral network at low values of $x$. The number of network breaking defects in the form of non-bridging oxygens (NBO's) increases slowly with $x$ until $x$=0.5 (NBO's ~ 10% at $x$=0.5). By $x$=0.67 the network breaking defects become significant as evidenced by the significant drop in the *average* coordination number of Si. By contrast, Al-O tetrahedra remain free of NBO's and fully integrated in the Al/Si-O network for all values of $x$. Calcium maintains a rather uniform coordination sphere of approximately 5 oxygen atoms for all values of $x$. The results suggest that not only Si/Al-O tetrahedra but Ca-O polyhedra, too, play a role in determining the glassy structure.

Silicate glasses in general, and aluminosilicate glasses in particular, are of great industrial and geological importance. Despite this, some very basic issues about their structure are still not well established. It is well known that silica glass is a continuous network of corner-shared Si-O tetrahedra [1]. At the vertex of each tetrahedron, and shared between two tetrahedra, is an oxygen atom known as a bridging oxygen (BO). When $Si^{4+}$ is partially replaced by another network former such as $Al^{3+}$ and, in addition, a modifier cation such as $Ca^{2+}$ is introduced to balance the charge, the glass structure changes. The change involves a gradual substitution of Al-O polyhedra for Si-O ones and an emerging of so-called non-bridging oxygens (NBO) linking one Si-O or Al-O polyhedral unit with a Ca-based one [2]. The presence of NBO's, i.e. the breaking of the network, has a large effect on the thermo-dynamics of the glass and on important properties such as the viscosity of the melt [2]. The exact nature of the modifier ion (Ca) coordination and of the way the NBOs are introduced into the network has been the subject of intensive research [3-7] yet is still not fully resolved. The coordination of Al, too, has been an issue for quite a time [4].

By utilizing the high fluxes of high energy x-rays available at third generation synchrotron x-ray sources we have obtained very high real-space resolution atomic pair distribution functions (PDF's) from a series of $Ca_{x/2}Al_xSi_{1-x}O_2$ glasses with x=0,0.25,0.5, 0.67. This has allowed us, for the first time, to resolve the different Si-O and Al-O bonds at r=1.60 Å and r=1.75 Å, respectively, and study the Si-O and Al-O coordination polyhedra independently. This has proved impossible in lower resolution structural studies up to now because Si and Al have similar scattering cross-sections to each other for both x-rays and neutrons.

The glasses we studied all have testosilicate compositions whereby the modifier cations exactly charge-balance the Al in the network and it is not *a priory* necessary for them to act as network breakers. In this case, it was thought that they occupy interstitial sites in a random way and do not participate in the structure formation [1]. Recent NMR results draw this simple view into question by finding spectroscopic evidence for the presence of 10 times more NBO's than expected from the nominal composition of $Ca_{0.25}Al_{0.5}Si_{0.5}O_2$ glass [7]. Our structural data on a similar composition sample are in quantitative agreement with this view. They also reveal some interesting facts about the distribution of NBO's in the network. It turns out that NBO's are located presumably on Si-O but not Al-O tetrahedra. This is evidenced by the fact that the *average* coordination of Al remains constant at approximately 4.0(1) while that of Si drops from 4.0(1) to 3.2(1) when $x$ increases from 0 to 0.67. Interestingly, the coordination environment of Ca ions is rather uniform and well defined regardless of the Si/Al ratio and the abundance of NBO's.

The atomic PDF, $G(r) = 4\pi r[\rho(r) - \rho_o]$, gives the deviation of the local, $\rho(r)$, from the average, $\rho_o$, atomic number-density and so describes the atomic arrangement in materials. It is the sine Fourier transform of the experimentally observable total structure factor, $S(Q)$, i.e.

$$G(r) = (2/\pi) \int_{Q=0}^{Q_{max}} Q[S(Q)-1] \sin(Qr) dQ, \quad (1)$$

where $Q$ is the magnitude of the wavevector. The structure factor is related to the elastic part of the total diffracted intensity, $I^{el.}(Q)$, as follows:

$$S(Q) = 1 + \left[ I^{el.}(Q) - \sum c_i |f_i(Q)|^2 \right] / \left| \sum c_i f_i(Q) \right|^2, \quad (2)$$

where $c_i$ and $f_i$ are the atomic concentration and scattering factor respectively for the atomic species of type $i$ [8]. Very high real-space resolution is required to resolve the distinct Si-O and Al-O coordination polyhedra in aluminosilicate glasses. High real-space resolution is obtained by measuring $I^{el.}(Q)$ to a very high value of Q ($Q_{max} \geq 40$ Å$^{-1}$) by using high-energy x rays. The experiments were carried out at the 1-ID beam line at the Advanced Photon Source (APS), Argonne. Four samples of $Ca_{x/2}Al_xSi_{1-x}O_2$ family with $x=0, 0.25, 0.5, 0.67$, all uniform plates of thickness ≤ 1 mm, were measured. The glasses were made by standard procedures [4]. The measurements were done in symmetric transmission geometry at 20 K. Low temperature was used to minimize thermal vibration in the samples, and hence to improve the resolution of the experimental PDFs. A bent double-Laue Si (111) crystal monochromator was used to monochromatize the white beam and deliver intense flux of photons of energy 80.6 keV. The use of x rays of such high an energy makes it possible for the higher wave vectors to be reached and reduces several unwanted experimental effects such as absorption and multiple scattering. Measuring $I^{el.}(Q)$ to very high-values of $Q$ is, however, not an easy task in the case of materials composed of light atomic species even when a third-generation synchrotron source is employed. The main difficulty stems from the fact that the elastic scattering is only a small fraction of the total diffracted intensity at high wave vectors as demonstrated in Fig. 1. This experimental difficulty was tackled in the following way: Diffracted x-ray photons were collected with an intrinsic germanium detector connected to a multi-channel analyzer. The elastic component of the diffracted x-ray intensity was extracted during data collection by setting proper energy windows. The procedure is straightforward at higher wave vectors where the elastic and inelastic (Compton) intensities are well separated from one another (see Fig. 1). The part of Compton scattering at low values of Q not eliminated by the preset energy window was removed analytically applying a procedure suggested by Ruland [9]. The procedure has already been successfully tested in a high-energy x-ray diffraction study on the local structure of In-Ga-As alloys [10]. Several diffraction runs were conducted with each sample, and the elastic intensities collected were averaged to improve the statistical accuracy and reduce any systematic effect effect due instabilities in the experimental set up. The data were normalized for flux, corrected for background scattering, detector dead time and absorption, and then converted to structure factors as defined by Eq. 2. The results are shown in Fig. 2. All data processing was done using the program RAD [11]. To further improve the statistical accuracy of the experimental data in reciprocal space and reduce the unphysical high-frequency noise in real space, a smooth line was placed through the Q[S(Q)-1] data points of wave vectors higher than 20 Å$^{-1}$ with the use of statistical procedures based on the Maximum Entropy Method [12]. The data for the Fourier transform were then terminated at $Q_{max}$= 40 Å$^{-1}$ beyond which the signal to noise ratio of the unsmoothed data became unfavorable. It should be noted that this is a very high wave vector for x-ray diffraction measurements on glassy materials. For comparison, $Q_{max}$ from a laboratory source is usually less than 20 Å$^{-1}$ [13].

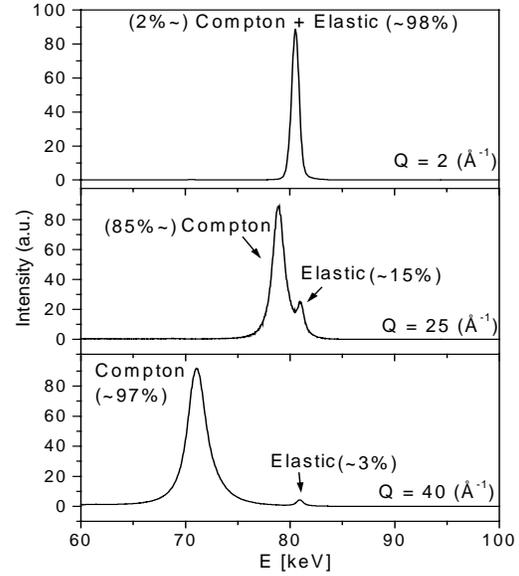

Figure 1. Total x-ray diffraction energy spectra from $Ca_{0.25}Al_{0.5}Si_{0.5}O_2$ glass collected at three different wave vectors. Data were taken with x-rays of energy 80.6 keV. Note the dramatic decrease of the elastic scattering with the increase of Q.

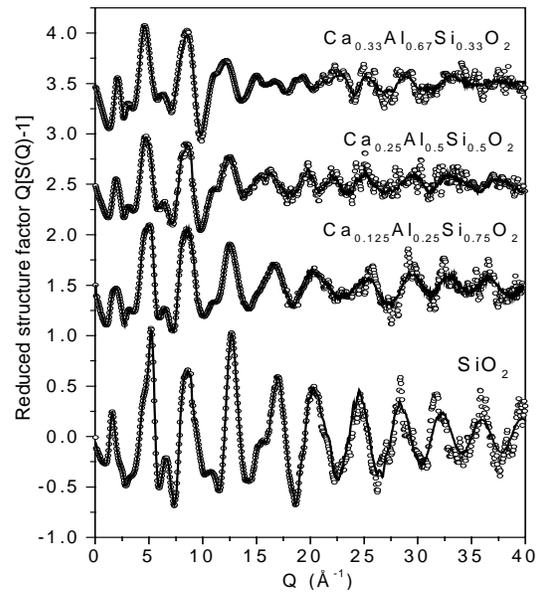

Figure 2. Reduced structure factors for calcium aluminosilicate glasses (symbols) together with the optimum smooth line.

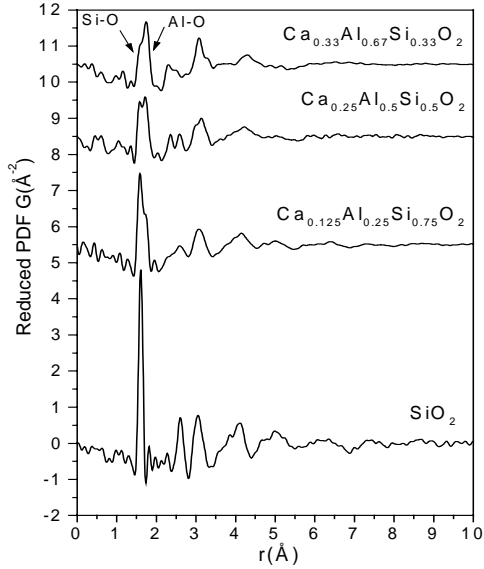

Figure 3. Reduced atomic PDFs for $Ca_{x/2}Al_xSi_{1-x}O_2$ glasses ($x$=0,0.25,0.5,0.67) obtained by Fourier transforming the smoothed data of Fig. 2.

As can be seen in Fig. 2 the experimental structure factors exhibit prominent oscillations up to the maximum Q value reached. These oscillations may only come from the presence of well-defined coordination polyhedra building the glassy structure. These polyhedra are seen as well defined peaks in the low-r region of the experimental PDFs shown in Fig. 3. The first peak in the PDF for $SiO_2$ glass is positioned at 1.61(1) Å and is very sharp. By fitting the peak with a Gaussian one finds that in silica glass the Si-O coordination number, Z, is 4.0(1) (see Fig. 4 and Table 1) and the root-mean-square deviation, $<u^2>^{1/2}$, from the average Si-O distance is 0.040(2) Å. The present results are in rather good agreement with those of Grimley et al. ( Si-O = 1.609(4) Å ; Z = 3.85(16) and $<u^2>^{1/2}$ =0.047(4) Å) obtained by time-of-flight neutron diffraction experiments [14].

As can be seen in figs. 3 and 4 the first peak in the experimental PDFs for $Ca_{x/2}Al_xSi_{1-x}O_2$ glasses (x=0.25, 0.5,0.67) is composed of two components. The first one, peaking at approximately 1.6 Å, becomes lower in intensity while the intensity of the second one, peaking at approximately 1.75 Å, increases with $x$. The corresponding first coordination numbers and distances determined by a Gaussian fit to the data, as shown in fig. 4, are summarized in Table 1. They can be unambiguously attributed to Si-O and Al-O tetrahedral units [2,13]. Thus the present high-energy x-ray diffraction experiments have yielded atomic PDFs with resolution high enough to resolve the two distinct Si-O and Al-O coordination polyhedra in aluminosilicate glasses.

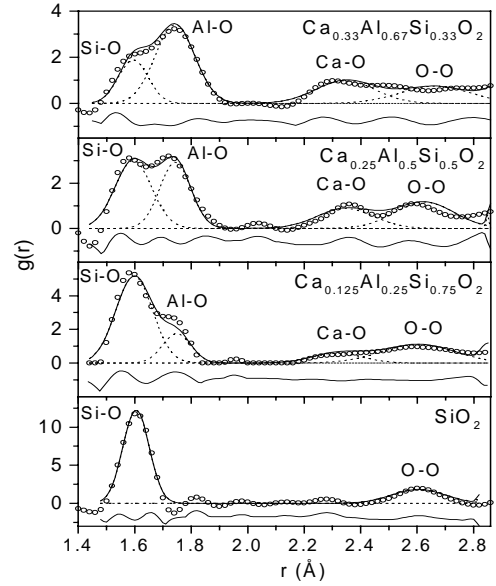

Figure 4. Gaussian fit to the first peaks in the PDFs, g(r) = ρ(r)/ρ$_o$, for $Ca_{x/2}Al_xSi_{1-x}O_2$ (x=0,0.25,0.5,0.67) glasses. Experimental data: symbols; fitted data: full line; individual gaussians: broken line; residual difference: full line (bottom). Peaks are labeled with the corresponding atomic pairs.

A careful analysis of the *average* Si-O and Al-O coordination numbers reveals the following characteristics of the atomic ordering in $Ca_{x/2}Al_xSi_{1-x}O_2$ glasses. Both Si and Al atoms are coordinated by four oxygens and thus participate in a continuous tetrahedral network when Al/Ca content is low ($x$=0, 0.25). By $x$=0.5 the network begins to break as the slight decrease in the *average* coordination number of Si indicates. This result agrees well with the findings of $^{17}$O quantum MAS-NMR experiments [7] and supports the view that NBO's can be present in aluminosilocate glasses even though it is not required by stoichiometry. When more that half of the Si ions are replaced by Al ($x$=0.67) the number of NBO's in the first coordination sphere of Si increases further making the *average* Si-O coordination number as low as 3.2(1). Al-O tetrahedra, however, remain free of NBOs. The result suggests that the disintegration of the tetrahedral network proceeds mainly by the creation of Si-O-Ca and not Al-O-Ca linkages.

TABLE 1. First neighbour Si-O, Al-O, Ca-O and O-O distances and *average* coordination numbers in $Ca_{x/2}Al_xSi_{1-x}O_2$ glasses (x=0,0.25,0.5,0.67) obtained by a Gaussian fit to the corresponding high-resolution PDFs, $g(r) = \rho(r)/\rho_o$.

| x | Si-O | Al-O | Ca-O |
|---|---|---|---|
| 0 | 1.61(1) Å / 4.0(1) | | |
| 0.25 | 1.60(1) Å / 3.95(10) | 1.75(1) Å / 3.95(10) | 2.32(2) Å / 5.3(2) |
| 0.5 | 1.60(1) Å / 3.85(10) | 1.75(1) Å / 4.0(1) | 2.36(2) Å / 5.2(2) |
| 0.67 | 1.60(1) Å / 3.2(1) | 1.75(1) Å / 3.95(10) | 2.34(2) Å / 5.3(2) |

This result does not imply that Si-O tetrahaedra are oxygen deficient, even at high calcium concentrations. We envisage the following situation. In a fully connected network every network-former Si and Al is four-fold coordinated and every oxygen has two network-former cations at well defined distances of 1.6 and 1.75 A from it. At high Ca content some of the oxygens become NBO's and change their coordination from having two network-former cation neighbors to having one network-former (Si) and one network-modifier neighbour (Ca) cation. In this way the emerging of an NBO effectively removes a Si-O bond without necessarily creating an oxygen vacancy as already discussed in [4,7]. This results in a relative loss in intensity in the Si-O peak in the PDF and a drop in the average Si-O coordination number although locally all Si are fully tetrahedrally coordinated. Thus the observed change in the average Si-O coordination numbers can be directly linked to the change in the connectivity of the tetrahedral network.

The disintegration of the tetrahedral network does not seem to affect the Ca-O first neighbour distribution seen as a peak positioned at approximately 2.3 Å in the experimental PDFs [3]. The inspection of the data given in Table 1 shows that the *average* oxygen coordination of Ca does not change even though the Si /Al ratio and the number of NBO's change significantly. Also, the rms deviation from the average Ca-O distance ($<u^2>^{1/2}$ (Ca-O) = 0.080(2) Å) is quite small (of the order of the broadening of the coordination shell of Al ($<u^2>^{1/2}$ (Al-O) = 0.054(2) Å)) for all values of x. The result implies that Ca has a well-defined and stable oxygen coordination. By demanding and acquiring it Ca atoms would inevitably affect the way the tetrahedral backbone of $Ca_{x/2}Al_xSi_{1-x}O_2$ glasses arranges in space.

In summary, the present work demonstrates that:
i) By employing high-energy x-ray diffraction structure factors extended to at least 40 Å$^{-1}$ can be obtained even for materials composed of weakly scattering, light atomic species.
ii) The coordination polyhedra of Si and Al in aluminosilicate glasses can be clearly differentiated and, hence, independently analyzed. In particular, $Ca_{x/2}Al_xSi_{1-x}O_2$ glasses (x=0.,0.25,0.5, 0.67) are built up of interconnected Si-O and Al-O tetrahedra with the degree of their connectivity decreasing with increasing *x*.
iii) In $Ca_{x/2}Al_xSi_{1-x}O_2$ glasses (x=0.,0.25,0.5,0.67) the breaking of Si/Al-O network proceeds via the creation of NBO's located on Si-O but not Al-O tetrahedra.
iv) Even when two strong network-formers such as Si and Al as well as NBO's are present the network-modifier Ca acquires a well-defined and constant oxygen coordination and so it is very likely to play a role in the formation of the glass structure.

We would like to thank I-K. Jeong for the help with the experiments. The work was supported by NSF grant CHE-9903706. The Advanced Photon Source is supported by DOE under contract W-31-109-Eng-38.